\documentclass{article}

\usepackage{arxiv}
\usepackage[utf8]{inputenc} 
\usepackage[T1]{fontenc}    
\usepackage{hyperref}       
\usepackage{url}            
\usepackage{booktabs}       
\usepackage{amsfonts}       
\usepackage{nicefrac}       
\usepackage{microtype}      
\usepackage{lipsum}
\usepackage{graphicx}
\usepackage{cite}
\usepackage{amsmath,amssymb,amsfonts}
\usepackage{algorithmic}
\usepackage{graphicx}
\usepackage{textcomp}
\usepackage{xcolor}
\usepackage{multirow}
\usepackage{booktabs}
\usepackage{array}
\usepackage{graphicx} 
\usepackage{amsmath}
\usepackage{subcaption}
\usepackage{csquotes}
\usepackage{enumitem}
\usepackage{hyperref}
\usepackage{cleveref}
\usepackage{balance}
\usepackage{fancyhdr}
\usepackage[ruled,vlined,linesnumbered]{algorithm2e}
\usepackage{xurl}
\usepackage{fancyhdr}
\graphicspath{ {./images/} }

\begin{document}

\thispagestyle{empty}
{\Huge \textbf{IEEE Copyright Notice}}\\[1em]
{\large
\noindent
\textcopyright~2024 IEEE. Personal use of this material is permitted. Permission from IEEE must be obtained for all other uses, in any current or future media, including reprinting/republishing this material for advertising or promotional purposes, creating new collective works, for resale or redistribution to servers or lists, or reuse of any copyrighted component of this work in other works.

DOI: \href{https://doi.org/10.1109/ICCIT64611.2024.11022605}{10.1109/ICCIT64611.2024.11022605}
}

\newpage

\fancypagestyle{titlepage}{
  \fancyhf{}
  \fancyhead[C]{\footnotesize This work has been accepted for publication in 2024 27th International Conference on Computer and Information Technology (ICCIT).\\
  The final published version is available via IEEE Xplore.\\
  DOI: \href{https://doi.org/10.1109/ICCIT64611.2024.11022605}{10.1109/ICCIT64611.2024.11022605}}
  \renewcommand{\headrulewidth}{0pt}
}

\title{Comprehensive Lung Disease Detection Using Deep Learning Models and Hybrid Chest X-ray Data with Explainable AI}

\author{
 Shuvashis Sarker \\
  Department of Computer Science and Engineering\\
  Ahsanullah University of Science and Technology\\
  Dhaka,Bangladesh\\
  \texttt{shuvashisofficial@gmail.com} \\
   \And
Shamim Rahim Refat \\
  Department of Computer Science and Engineering\\
  Ahsanullah University of Science and Technology\\
  Dhaka,Bangladesh\\
  \texttt{n.a.refat2000@gmail.com} \\
  \And
 Faika Fairuj Preotee \\
  Department of Computer Science and Engineering\\
  Ahsanullah University of Science and Technology\\
  Dhaka,Bangladesh\\
  \texttt{faikafairuj2001@gmail.com} \\
 \And
  Tanvir Rouf Shawon \\
  Department of Computer Science and Engineering\\
  Ahsanullah University of Science and Technology\\
  Dhaka,Bangladesh\\
  \texttt{shawontanvir95@gmail.com} \\
  \And
 Raihan Tanvir \\
  Department of Computer Science and Engineering\\
  Ahsanullah University of Science and Technology\\Dhaka,Bangladesh\\
  \texttt{raihantanvir@gmail.com}\\
}

\maketitle
\thispagestyle{titlepage}

\begin{abstract}
Advanced diagnostic instruments are crucial for the
accurate detection and treatment of lung diseases, which affect
millions of individuals globally. This study examines the effectiveness of deep learning and transfer learning models using a
hybrid dataset, created by merging four individual datasets from
Bangladesh and global sources. The hybrid dataset significantly
enhances model accuracy and generalizability, particularly in
detecting COVID-19, pneumonia, lung opacity, and normal lung
conditions from chest X-ray images. A range of models, including
CNN, VGG16, VGG19, InceptionV3, Xception, ResNet50V2, InceptionResNetV2, MobileNetV2, and DenseNet121, were applied
to both individual and hybrid datasets. The results showed superior performance on the hybrid dataset, with VGG16, Xception,
ResNet50V2, and DenseNet121 each achieving an accuracy of
99\%. This consistent performance across the hybrid dataset
highlights the robustness of these models in handling diverse
data while maintaining high accuracy. To understand the models’
implicit behavior, explainable AI techniques were employed to
illuminate their black-box nature. Specifically, LIME was used
to enhance the interpretability of model predictions, especially
in cases of misclassification, contributing to the development
of reliable and interpretable AI-driven solutions for medical
imaging.
\end{abstract}


\section{Introduction}
Palmo aka lungs often referred to as the \textit{silent powerhouses} of the body, play a vital role in sustaining life by ensuring the delivery of essential oxygen to every cell and the removal of carbon dioxide. Lung diseases, encompassing a broad spectrum of conditions, can range from mild afflictions to severe, life-threatening disorders. The advent of the COVID-19 pandemic, caused by the SARS-CoV-2 virus, has brought unprecedented attention to the significance of respiratory health, as the virus primarily targets the respiratory system.
The rapid global spread of COVID-19 transformed it into a pandemic, with the World Health Organization (WHO) reporting an overwhelming 775,867,547 cases and 7,057,145 deaths worldwide as of August 4, 2024\footnote{\url{https://data.who.int/dashboards/covid19/cases?n=}}. 
The DGHS found 2,051,190 cases and 29,499 deaths in Bangladesh. The age group 15–34 had the most cases, and the age group 60 and older had the most deaths\footnote{\url{https://dashboard.dghs.gov.bd/pages/covid19}}.

The complexities of pulmonary diseases across diverse populations and the necessity of a more comprehensive approach to appreciating their impact are recognized in our investigation. The ability to apply of findings across broader demographics may be restricted by the fact that existing research has frequently concentrated on region-specific datasets. In order to resolve this matter, we have combined datasets from various regions, including two from Bangladesh and two from other regions, into a singular hybrid dataset. This method allows for a comprehensive analysis of the impact of increased data diversity on the efficacy of deep learning models in the context of lung diseases. The objectives of our study are threefold:
\begin{itemize}
    \item[i.] To evaluate the performance of Deep learning and Transfer learning models on individual Regional Datasets.
    \item[ii.] To assess how these models perform when applied to a Hybrid Dataset created from multiple regional datasets, including those from Bangladesh and global sources.
    \item[iii.] To utilize Explainable AI techniques to identify and understand the reasons behind misclassifications, thereby enhancing the interpretability of the results.
\end{itemize}
 The opportunities and challenges of utilizing a variety of datasets to analyze pulmonary disease are the focus of our research. This method improves our understanding of the efficacy of models in various contexts.
 The remainder of this paper is structured as follows: Section \ref{Related Work} discusses the related work, while Section \ref{Methodology} outlines the proposed methodology, and Section \ref{Result Analysis} presents the analysis of the results. Techniques related to Explainable AI are detailed in Section \ref{Explainable AI}, Limitations and Future Works are described in Section \ref{Limitations and Future Work} and concluding remarks are provided in Section \ref{Conclusion}

\section{Related Work}
\label{Related Work}
\subsection{CNN and Pre-trained Models}

Talukder et al. (2024)\cite{talukder2024empowering} utilized two datasets, the COVID-19 X-ray dataset and the Chest X-ray Image Dataset \cite{https://doi.org/10.17632/m4s2jn3csb.1}, to fine-tune several deep learning models for COVID-19 and lung disease detection. They tested models like Xception, InceptionResNetV2, ResNet50, ResNet50V2, EfficientNetB0, and EfficientNetB4, achieving up to 100\% accuracy in COVID-19 detection with EfficientNetB4. Additionally, EfficientNetB4 also demonstrated high accuracy (99.17\%) in lung disease detection, highlighting the effectiveness of fine-tuned transfer learning models for medical imaging tasks. Similarly,Hariri et al (2023)\cite{hariri2023covid} created a simple and efficient CNN model to differentiate between COVID-19, viral pneumonia, and bacterial pneumonia using chest X-rays. Their model achieved an accuracy of 89.89\%, performing better than other popular models like EfficientNet B2, which had an accuracy of 85.7\%. At the same time, Hasan et al. (2023)\cite{hasan2023deep} applied deep transfer learning using VGG16 and ResNet50 models to classify COVID-19 and pneumonia in chest X-ray images. They fine-tuned these pre-trained models on a dataset of 9218 images, achieving accuracies of 89.23\% and 88.80\%, respectively. In 2023 study, Saleh et al.\cite{saleh2023comparative} also analyzed AI models for predicting COVID-19 using a meticulously curated hybrid dataset \cite{https://doi.org/10.17632/9xkhgts2s6.4} of 36,170 chest X-ray images. The dataset addressed disparities in histogram characteristics, patient demographics, and scanning environments, ensuring consistency through normalization and bi-cubic interpolation. This diverse dataset enabled models like DenseNet121, InceptionResNetV2, ResNet152V2, VGG16, and Xception—particularly ResNet152V2—to achieve high accuracy (96.17\%) and strong generalization across different populations, highlighting the value of a well-prepared hybrid dataset in medical image analysis. Additionally, Roy et al. (2022)\cite{roy2022svd} evaluated several CNN models, including ResNet-50, InceptionV3, Xception, DenseNet-121, VGG-16, and VGG-19, for detecting COVID-19 from chest X-ray images. They addressed class imbalance in the highly imbalanced CXR dataset by using SVD-CLAHE Boosting and a BWCCE loss function. The ResNet-50 model, with these enhancements, outperformed the others, achieving 94\% accuracy and a 96\% AUC. Sunyoto et al.\cite{sunyoto2022performance} evaluated the performance of the VGG16 model using nine publicly available chest X-ray datasets in 2022, comprising 38,181 images across four categories: normal, pneumonia, viral pneumonia, and COVID-19. The model achieved a maximum accuracy of 97.99\%, demonstrating robust performance. Srivastava et al.introduced CoviXNet in 2022, a lightweight CNN for COVID-19 detection using chest X-rays\cite{srivastava2022covixnet}. Comparing it with models like InceptionV3, Modified EfficientNet B0 \& B1, and ResNetV2 on 6207 X-rays, CoviXNet achieved 99.47\% accuracy in binary classification and 96.61\% in 3-class classification, offering strong performance with fewer parameters. InceptionV3 and Modified EfficientNet slightly outperformed in binary classification (99.78\% accuracy) but required more computational resources.

\subsection{Explainable Artificial Intelligence}
Ifty et al.\cite{ifty2024explainable} did a study in 2024 that looked at how well different deep learning models could grade lung diseases from chest X-rays. Using stratified 5-fold cross-validation, the Xception model got the best results, with a 96.21\% success rate. It did better than DenseNet121 and InceptionV3, which both got to about 92\%. The study also looked at combination models, and the InceptionV3+Xception hybrid did very well. XAI methods like Grad-CAM and LIME were used to make the explanations more visual. Mahamud et al. (2024)\cite{mahamud2024explainable} developed a Dense CNN with Modified Dense Blocks and Transition Layers to classify lung diseases like COVID-19, pneumonia, and tuberculosis. Using transfer learning and data augmentation, their model achieved 99.20\% accuracy with 99\% precision and recall, outperforming EfficientNetB0, InceptionV3, and LeNet. The study also employed Explainable AI techniques, including SHAP, LIME, Grad-CAM, and Grad-CAM++, to enhance model interpretability. Gasca Cervantes et al.\cite{cervantes2021lime} used 10,686 chest X-rays to test four CNN models that had already been trained: VGG16, DenseNet201, ResNet50, and EfficientNetB3. The most accurate network was DenseNet201, with 96.47\% of the time. It was followed by VGG16 (95.71\%), ResNet50 (93.15\%), and EfficientNetB3 (92.39\%). DenseNet201 did the best, but VGG16 didn't focus on useful lung regions when LIME-based XAI analysis was done. Ong et al. (2021)\cite{ong2021comparative} further utilized the SqueezeNet to classify chest X-rays into the categories of COVID-19, normal, and pneumonia with an 84.3\% accuracy on the COVIDx dataset. They used XAI techniques, LIME and SHAP, for interpretability, where SHAP gives more consistent and accurate explanations in model understanding through the importance of the highlighted critical regions of lungs for an improved interpretation in medical imaging.

\section{Methodology}
\label{Methodology}
This research aims to develop a system to identify COVID-19, pneumonia, lung opacity, and normal lung conditions using chest X-ray images. \Cref{Figure 2} illustrates the methodology of our study which involves collecting a comprehensive dataset, preprocessing iteratively, fine-tuning hyperparameters, creating a hybrid dataset for data analysis, and implementing a LIME explainer on the best-performing model to enhance transparency and trustworthiness.
\begin{figure}[!h]
    \centering
    \includegraphics[width=\linewidth]{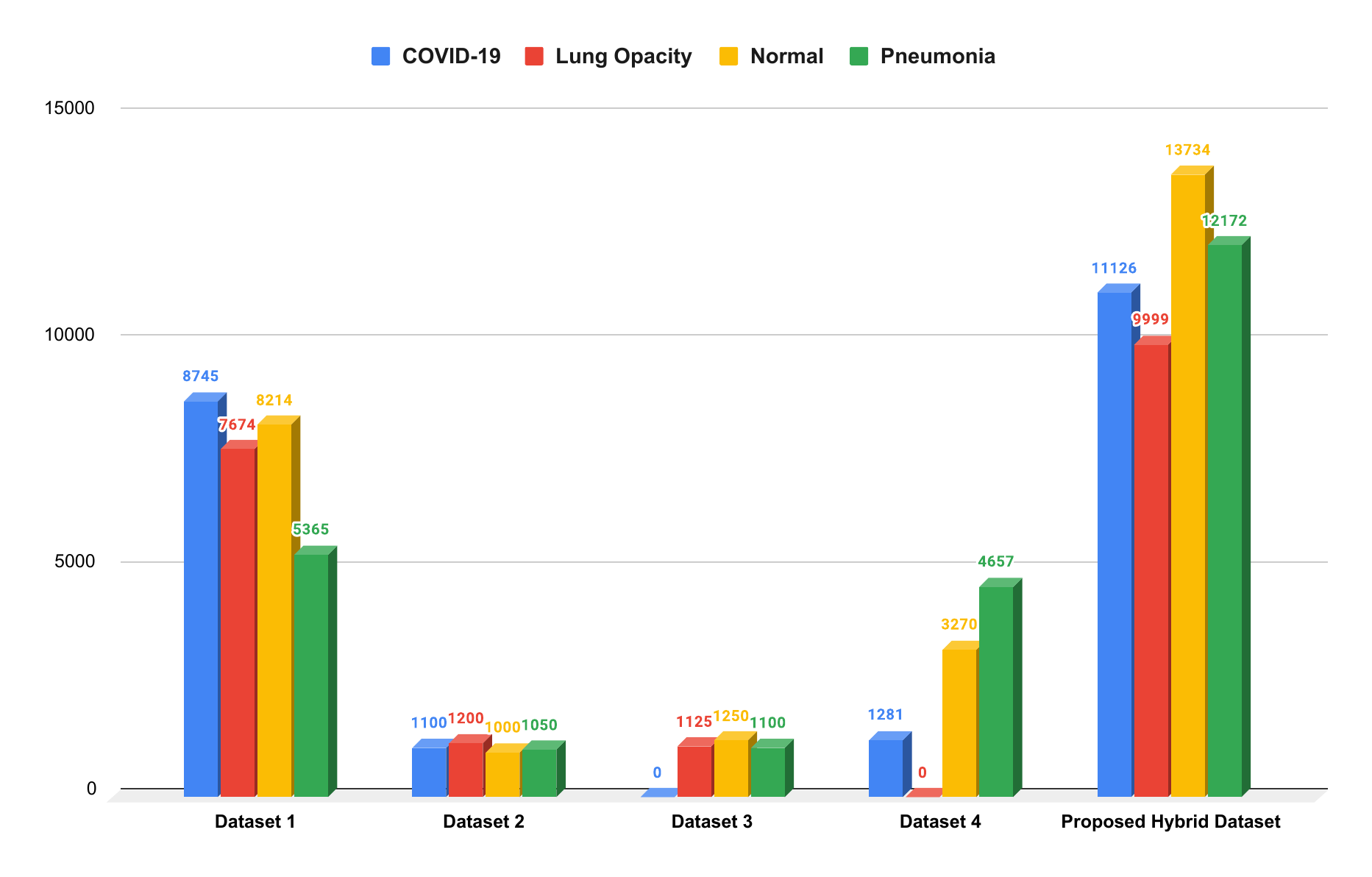}
    \caption{Comparison of Different Datasets by Categories}
    \label{Figure 1}
\end{figure}

\subsection{Dataset}
In our study, we have utilized four publicly available datasets to enhance the detection capabilities of lung conditions through Chest X-ray images. \textbf{Dataset 1}, \textit{Balanced Augmented Covid CXR Dataset}, developed by Mrinal Tyagi \cite{MrinalTyagi} and published on Kaggle, is an improved version of the widely recognized COVID-19 Radiography Database. This dataset, containing a total of 29,998 images, was balanced using various techniques to address significant class imbalance issues. \textbf{Dataset 2}, \textit{Chest X-Ray Image Dataset}, compiled by Md Alamin Talukdar \cite{https://doi.org/10.17632/m4s2jn3csb.1} and available on Mendeley Data, includes 4,350 images were collected from hospitals in Bangladesh, meticulously categorized into four classes to facilitate research and diagnostic efforts related to lung conditions. \textbf{Dataset 3}, \textit{Lung X-Ray Image Dataset}, also assembled by Md Alamin Talukdar \cite{https://doi.org/10.17632/9d55cttn5h.1} and available on Mendeley Data, comprised 3,475 X-ray images classified into three distinct categories, gathered from various healthcare sources to support the detection and diagnosis of lung conditions. Lastly, \textbf{Dataset 4}, \textit{Curated Dataset for COVID-19 Posterior-Anterior Chest Radiography Images (X-Rays)}, was prepared by Unais Sait et al \cite{https://doi.org/10.17632/9xkhgts2s6.4}. and publicly available on Mendeley, is a comprehensive collection obtained by aggregating 15 publicly available datasets, providing a diverse range of COVID-19 related X-ray images with a total of 4,350 images. In this dataset, the two classes \textbf{\textit{Viral Pneumonia}} and \textbf{\textit{Bacterial Pneumonia}} are merged into a single general class named \textbf{\textit{Pneumonia}} to maintain consistency across the datasets. To ensure uniformity, the class name \textbf{\textit{Viral Pneumonia}} in other datasets has also changed to \textbf{\textit{Pneumonia.}}

A proposed \textbf{hybrid dataset} which have been used in our study is a combined collection of these multiple datasets, integrated to create a more comprehensive and diverse data source for analysis and training purposes. \Cref{Figure 1} summarizes the distribution of images across different categories in the four datasets, highlighting the variety and volume of data available for each condition.


\subsection{Data Preprocessing}
The dataset we use containing X-ray images with varying shapes and resolutions. To standardize the input data and ensure consistency across the entire dataset,initially, all images from Dataset 2 and Dataset 3 were converted to grayscale for uniform analysis.After that, the images from Dataset 1 and the Hybrid Dataset were resized to 128x128 pixels, while the images from Datasets 2, 3, and 4 were standardized to 256x256 pixels. To prepare the data for model training, each dataset, including the Hybrid Dataset, was split into an 80:10:10 ratio using the Keras Utils library. In this split, 80\% of the images were allocated for training, 10\% for validation, and 10\% for testing.

\subsection{Proposed Method}
The proposed model architecture integrates a range of deep convolutional neural networks (CNNs), including VGG16, VGG19, ResNet50V2, InceptionV3, Xception, MobileNetV2, InceptionResNetV2, and DenseNet121, which are renowned for their image classification capabilities. VGG16 and VGG19, based on 3x3 convolutions, are widely used for feature extraction, with VGG19 having additional layers to capture more complex patterns. ResNet50V2 leverages residual connections to avoid vanishing gradients and improve training efficiency. InceptionV3 employs multi-scale feature extraction via various filter sizes (1x1, 3x3, 5x5) and reduction modules to minimize complexity. Xception, a variant of Inception, applies depthwise separable convolutions to improve efficiency. InceptionResNetV2 merges Inception modules with residual connections for enhanced feature learning and reduced vanishing gradients. MobileNetV2 is optimized for resource-constrained devices, utilizing inverted residuals and bottlenecks to balance performance and computational efficiency. Lastly, DenseNet121 ensures strong feature propagation by densely connecting each layer to all subsequent layers, thus mitigating the vanishing-gradient problem and reducing parameters \cite{salehi2023study}.

\begin{figure}[!h]
    \centering
    \includegraphics[width=\textwidth]{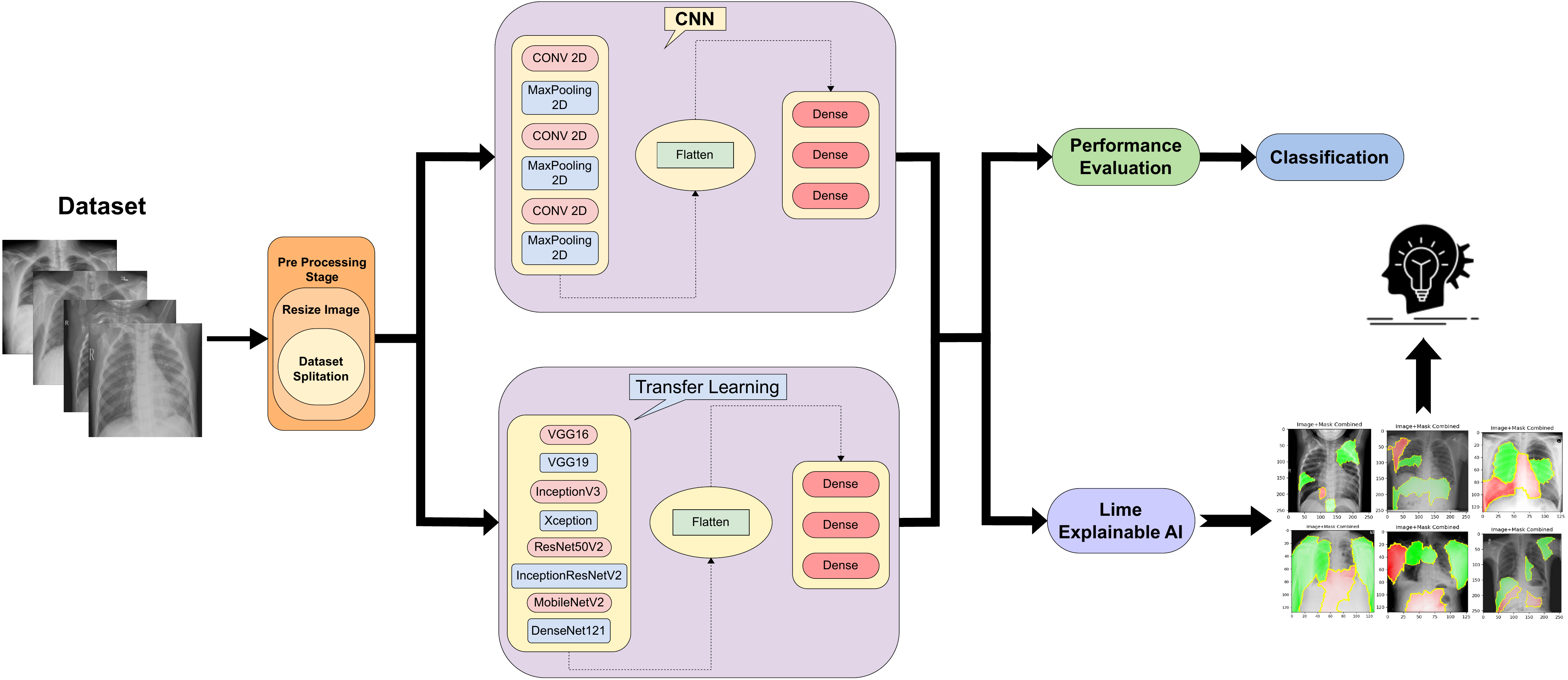}
    \caption{A Flow Diagram of Our Proposed Model}
    \label{Figure 2}
\end{figure}

In our study, we employed both CNNs and pre-trained models on four distinct datasets, as well as a hybrid dataset created by merging the four. Our CNN architecture consists of three Conv2D layers, increasing from 32 to 128 filters, and MaxPooling2D layers for dimensionality reduction. After flattening the data, two dense layers with 1024 and 512 neurons are applied, followed by dropout layers to reduce overfitting. The final softmax layer outputs four neurons corresponding to the dataset classes. For pre-trained models, we retained their base architecture and added custom fully connected layers (1024 and 512 neurons), along with dropout layers (0.5 and 0.25 dropout rates). Fine-tuning these pre-trained networks with our dataset allowed us to harness their feature extraction capabilities, improving performance while reducing training time.

\section{Result Analysis}
\label{Result Analysis}
The methodologies are executed on a computer system, with the best results selected for analysis. The system is equipped with an Intel Core i5 13400F processor, an NVIDIA GeForce RTX 3060 12GB GPU, and 16GB of DDR4 RAM. The models are implemented using Keras \footnote{\url{https://keras.io/}}, a high-level neural networks API, and TensorFlow \footnote{\url{https://www.tensorflow.org/}} as the backend. To ensure robustness, each model runs five times, and the best outcomes are recorded.

\begin{table}[htbp]
\centering
\caption{Performance metrics for different models across datasets}
\label{Table: Combined}
\renewcommand{\arraystretch}{1.2}
\begin{tabular}{ | c | c | c | c | c | c | }
\hline
\textbf{Dataset} & \textbf{Model} & \textbf{Accuracy} & \textbf{Precision} & \textbf{Recall} & \textbf{F1 Score} \\ \hline
\multirow{9}{*}{Dataset 1\cite{MrinalTyagi}} & CNN & 0.97 & 0.98 & 0.98 & 0.98 \\ \cline{2-6} 
 & \textbf{VGG16} & \textbf{0.99} & \textbf{0.99} & \textbf{0.99} & \textbf{0.99} \\ \cline{2-6} 
 & \textbf{VGG19} & \textbf{0.99} & \textbf{0.99} & \textbf{0.99} & \textbf{0.99} \\ \cline{2-6} 
 & \textbf{InceptionV3} & \textbf{0.99} & \textbf{0.99} & \textbf{0.99} & \textbf{0.99} \\ \cline{2-6} 
 & \textbf{Xception} & \textbf{0.99} & \textbf{0.99} & \textbf{0.99} & \textbf{0.99} \\ \cline{2-6} 
 & \textbf{ResNet50V2} & \textbf{0.99} & \textbf{0.99} & \textbf{0.99} & \textbf{0.99} \\ \cline{2-6} 
 & \textbf{InceptionResNetV2} & \textbf{0.99} & \textbf{0.99} & \textbf{0.99} & \textbf{0.99} \\ \cline{2-6} 
 & \textbf{MobileNetV2} & \textbf{0.99} & \textbf{0.99} & \textbf{0.99} & \textbf{0.99} \\ \cline{2-6} 
 & \textbf{DenseNet121} & \textbf{0.99} & \textbf{0.99} & \textbf{0.99} & \textbf{0.99} \\ \hline
\multirow{9}{*}{Dataset 2 \cite{https://doi.org/10.17632/m4s2jn3csb.1}} 
 & CNN & 0.84 & 0.85 & 0.84 & 0.84 \\ \cline{2-6} 
 & VGG16 & 0.93 & 0.93 & 0.93 & 0.93 \\ \cline{2-6} 
 & VGG19 & 0.91 & 0.91 & 0.91 & 0.91 \\ \cline{2-6} 
 & InceptionV3 & 0.89 & 0.89 & 0.89 & 0.89 \\ \cline{2-6} 
 & Xception & 0.91 & 0.91 & 0.91 & 0.91 \\ \cline{2-6} 
 & ResNet50V2 & 0.90 & 0.90 & 0.90 & 0.90 \\ \cline{2-6} 
 & InceptionResNetV2 & 0.87 & 0.87 & 0.87 & 0.87 \\ \cline{2-6} 
 & MobileNetV2 & 0.90 & 0.90 & 0.90 & 0.90 \\ \cline{2-6} 
 & \textbf{DenseNet121} & \textbf{0.94} & \textbf{0.94} & \textbf{0.94} & \textbf{0.94} \\ \hline
\multirow{9}{*}{Dataset 3 \cite{https://doi.org/10.17632/9d55cttn5h.1}} 
 & CNN & 0.91 & 0.93 & 0.91 & 0.92 \\ \cline{2-6} 
 & VGG16 & 0.93 & 0.93 & 0.93 & 0.93 \\ \cline{2-6} 
 & VGG19 & 0.90 & 0.90 & 0.90 & 0.90 \\ \cline{2-6} 
 & InceptionV3 & 0.92 & 0.92 & 0.92 & 0.92 \\ \cline{2-6} 
 & Xception & 0.91 & 0.91 & 0.91 & 0.91 \\ \cline{2-6} 
 & ResNet50V2 & 0.92 & 0.92 & 0.92 & 0.92 \\ \cline{2-6} 
 & InceptionResNetV2 & 0.90 & 0.90 & 0.90 & 0.90 \\ \cline{2-6} 
 & MobileNetV2 & 0.91 & 0.92 & 0.91 & 0.91 \\ \cline{2-6} 
 & \textbf{DenseNet121} & \textbf{0.94} & \textbf{0.94} & \textbf{0.94} & \textbf{0.94} \\ \hline
\multirow{9}{*}{Dataset 4 \cite{https://doi.org/10.17632/9xkhgts2s6.4}} 
 & CNN & 0.97 & 0.98 & 0.96 & 0.97 \\ \cline{2-6} 
 & \textbf{VGG16} & \textbf{0.98} & \textbf{0.99} & \textbf{0.98} & \textbf{0.99} \\ \cline{2-6} 
 & \textbf{VGG19} & \textbf{0.98} & \textbf{0.99} & \textbf{0.99} & \textbf{0.99} \\ \cline{2-6} 
 & \textbf{InceptionV3} & \textbf{0.98} & \textbf{0.98} & \textbf{0.98} & \textbf{0.98} \\ \cline{2-6} 
 & \textbf{Xception} & \textbf{0.98} & \textbf{0.99} & \textbf{0.98} & \textbf{0.99} \\ \cline{2-6} 
 & \textbf{ResNet50V2} & \textbf{0.98} & \textbf{0.98} & \textbf{0.98} & \textbf{0.98} \\ \cline{2-6} 
 & InceptionResNetV2 & 0.97 & 0.98 & 0.97 & 0.97 \\ \cline{2-6} 
 & \textbf{MobileNetV2} & \textbf{0.98} & \textbf{0.98} & \textbf{0.98} & \textbf{0.98} \\ \cline{2-6} 
 & \textbf{DenseNet121} & \textbf{0.98} & \textbf{0.98} & \textbf{0.98} & \textbf{0.98} \\ \hline
\multirow{9}{*}{\shortstack{Proposed \\ Hybrid \\ Dataset}}
 & CNN & 0.96 & 0.96 & 0.95 & 0.96 \\ \cline{2-6} 
 & \textbf{VGG16} & \textbf{0.99} & \textbf{0.99} & \textbf{0.99} & \textbf{0.99} \\ \cline{2-6} 
 & VGG19 & 0.98 & \textbf{0.99} & 0.98 & \textbf{0.99} \\ \cline{2-6} 
 & InceptionV3 & 0.98 & 0.98 & 0.98 & 0.98 \\ \cline{2-6} 
 & \textbf{Xception} & \textbf{0.99} & \textbf{0.99} & \textbf{0.99} & \textbf{0.99} \\ \cline{2-6} 
 & \textbf{ResNet50V2} & \textbf{0.99} & \textbf{0.99} & \textbf{0.99} & \textbf{0.99} \\ \cline{2-6} 
 & InceptionResNetV2 & 0.98 & 0.98 & 0.97 & 0.98 \\ \cline{2-6} 
 & MobileNetV2 & 0.98 & 0.98 & 0.98 & 0.98 \\ \cline{2-6} 
 & \textbf{DenseNet121} & \textbf{0.99} & \textbf{0.99} & \textbf{0.99} & \textbf{0.99} \\ \hline
\end{tabular}
\end{table}

The \Cref{Table: Combined} presents the performance metrics for various models across different datasets. For the individual datasets, Dataset 1 \cite{MrinalTyagi} shows exceptionally high accuracy across all models, with scores consistently reaching \textbf{99\%} for advanced models such as \textbf{VGG16}, \textbf{VGG19}, \textbf{InceptionV3}, \textbf{Xception}, \textbf{ResNet50V2}, \textbf{InceptionResNetV2}, \textbf{MobileNetV2}, and \textbf{DenseNet121}. This indicates that these models are highly effective when applied to Dataset 1 \cite{MrinalTyagi}. In contrast, Dataset 2 \cite{https://doi.org/10.17632/m4s2jn3csb.1} shows more variability, with \textbf{CNN} achieving an accuracy of \textbf{84\%}, while \textbf{DenseNet121} achieves the highest accuracy of \textbf{94\%}. Similarly, Dataset 3 \cite{https://doi.org/10.17632/9d55cttn5h.1} results range from \textbf{91\%} accuracy with \textbf{CNN} to \textbf{94\%} with \textbf{DenseNet121}. Dataset 4 \cite{https://doi.org/10.17632/9xkhgts2s6.4} also performs strongly, with \textbf{VGG16}, \textbf{VGG19}, and several other models consistently achieving accuracy around \textbf{98\%} to \textbf{99\%}.

For the hybrid dataset, the proposed combination of datasets maintains high performance, with most models achieving near-perfect accuracy. The \textbf{VGG16}, \textbf{Xception}, \textbf{ResNet50V2}, and \textbf{DenseNet121} models stand out, each reaching an accuracy of \textbf{99\%}, showcasing their robustness and ability to generalize well when faced with a more diverse dataset. This high consistency across the hybrid dataset demonstrates the effectiveness of these models in handling varied data while maintaining top-tier accuracy.
\begin{table}[htbp]
\centering
\caption{Comparison of Our Proposed Approach with Existing Methods}
\label{Table-3}
\renewcommand{\arraystretch}{1.1}
\begin{tabular}{|c|c|c|c|}
\hline
\textbf{Author Name} & \textbf{Approach} & \textbf{Accuracy} & \textbf{Using of XAI} \\ \hline
Talukder et al.\cite{talukder2024empowering} & \begin{tabular}[c]{@{}c@{}}EfficientNetB4 \\ + Fine-Tuning\end{tabular} & 99.17\% & NO \\ \hline
Roy et al \cite{roy2022svd}. & \begin{tabular}[c]{@{}c@{}}ResNet-50+\\ SVD-CLAHE\\ Boosting+\\ BWCCE\end{tabular} & 94\% & NO \\ \hline
Hariri et al \cite{hariri2023covid}. & CNN & 89.89\% & NO \\ \hline
Sunyoto et al. \cite{sunyoto2022performance} & VGG16 & 97.99\% & NO \\ \hline
Hasan et al \cite{hasan2023deep}. & VGG16 & 89.23\% & NO \\ \hline
Srivastava et al. \cite{srivastava2022covixnet} & CoviXNet & 99.47\% & NO \\ \hline
Saleh et al \cite{saleh2023comparative}. & ResNet50V2 & 96.17\% & NO \\ \hline
Ong et al \cite{ong2021comparative}. & SqueezeNet & 84.3\% & \begin{tabular}[c]{@{}c@{}}YES\\ (LIME\\ \& SHAP)\end{tabular} \\ \hline
Gasca Cervantes et al. \cite{cervantes2021lime} & DenseNet201 & 96.47\% & \begin{tabular}[c]{@{}c@{}}YES\\ (LIME)\end{tabular} \\ \hline
Ifty et al. \cite{ifty2024explainable} & \begin{tabular}[c]{@{}c@{}}Xception+\\ Fine-Tuning\end{tabular} & 96.21\% & \begin{tabular}[c]{@{}c@{}}Yes\\ (Grad-CAM\\ \& LIME)\end{tabular} \\ \hline
Mahamud et al. \cite{mahamud2024explainable} & Dense CNN & 99.20\% & \begin{tabular}[c]{@{}c@{}}YES\\ (LIME,SHAPE,\\ Grade-CAM,\\ Grade-Cam++\end{tabular} \\ \hline
\textbf{Proposed Model} & \textbf{\begin{tabular}[c]{@{}c@{}}Hybrid Dataset+\\ VGG16,Xception,\\ ResNet50V2,\\ DenseNet121\end{tabular}} & \textbf{99\%} & \textbf{\begin{tabular}[c]{@{}c@{}}YES\\ (LIME)\end{tabular}} \\ \hline
\end{tabular}
\end{table}
Overall, while all models have performed well across both individual and hybrid datasets, the advanced models such as \textbf{VGG16}, \textbf{Xception}, \textbf{ResNet50V2}, and \textbf{DenseNet121} have consistently showed superior accuracy, especially with the hybrid dataset. \Cref{Table-3} shows that our proposed model, utilizing a hybrid dataset with multiple CNN architectures and XAI techniques, achieves competitive accuracy compared to existing methods, many of which do not incorporate XAI. This consistency demonstrates that these models are highly effective at generalizing across diverse datasets, achieving the near-perfect accuracy we aimed for. The results confirm that our approach of combining multiple datasets into a hybrid dataset to test model performance was successful, validating that these models can handle varied data sources while maintaining high accuracy. This indicates that these models can work effectively for data from any region, making them versatile tools for broader applications.

\section{Explainable AI}
\label{Explainable AI}
Explainable Artificial Intelligence (XAI) seeks to enhance the transparency of AI models by providing clear explanations of their decision-making processes, which aids in improving prediction accuracy and building trust between humans and AI. XAI helps users understand the strengths and weaknesses of AI models, leading to a better comprehension of their behavior.
\subsection{LIME:}
LIME (Local Interpretable Model-agnostic Explanations) is an Explainable AI technique developed by Ribeiro et al.\cite{ribeiro2016should} to clarify how complex classifiers and regressors make predictions. LIME operates by locally approximating any model to a simplified, interpretable counterpart, thereby emphasizing the relationship between input features and outputs. It is model-agnostic, which means that it can consider the original model as a black box, rendering it versatile across a variety of data types, including images, text, and tabular data\cite{vimbi2024interpreting}. LIME functions by executing four fundamental procedures: permuting input data to generate artificial data points, predicting classes for these points, calculating their similarity to the original data, and employing a linear classifier to identify the most significant features that influence the model's decisions. This method enhances the transparency and comprehensibility of AI models\cite{maliyat2023investigating}.

\begin{figure}[htbp]
    \centering
    \begin{subfigure}[b]{0.48\textwidth}
        \includegraphics[width=\linewidth]{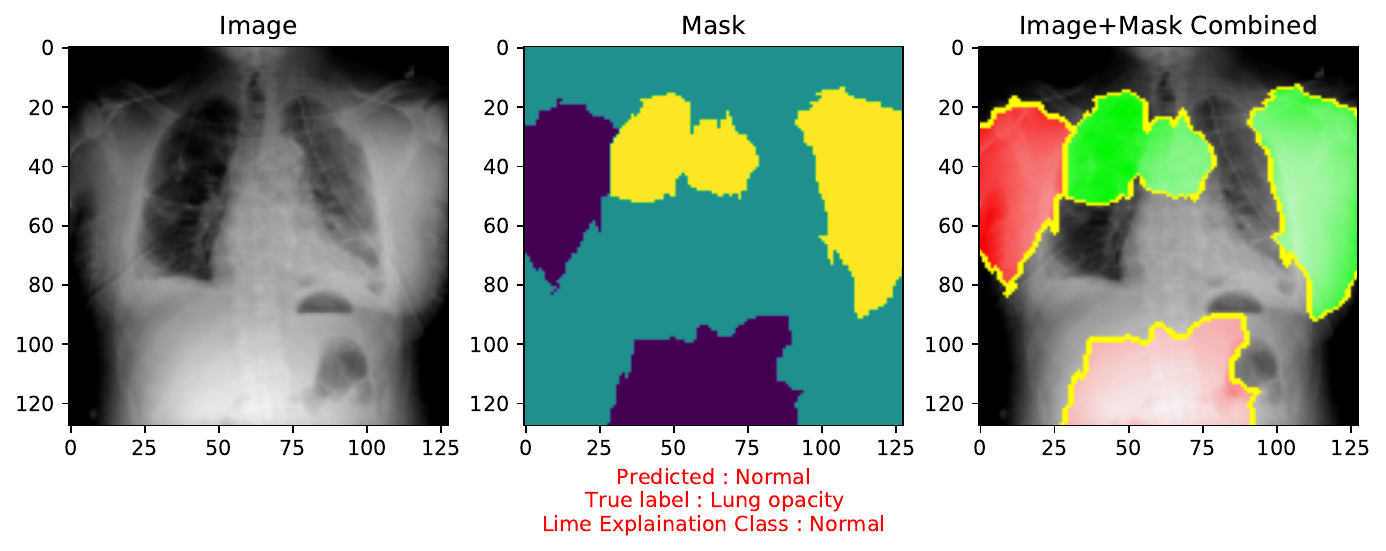}
        \caption{Misclassification Image of Dataset 1}
        \label{Figure 7(a)}
    \end{subfigure}
    \hfill
    \begin{subfigure}[b]{0.48\textwidth}
        \includegraphics[width=\linewidth]{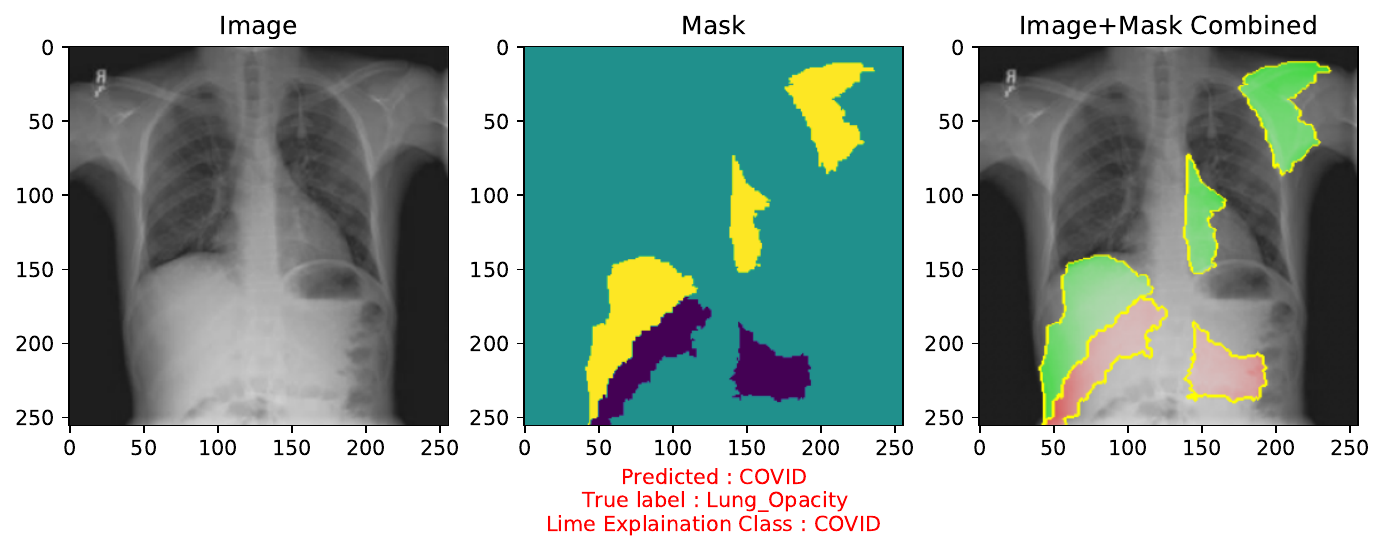}
        \caption{Misclassification Image of Dataset 2}
        \label{Figure 7(b)}
    \end{subfigure}
    
    \vspace{0.5em}
    
    \begin{subfigure}[b]{0.48\textwidth}
        \includegraphics[width=\linewidth]{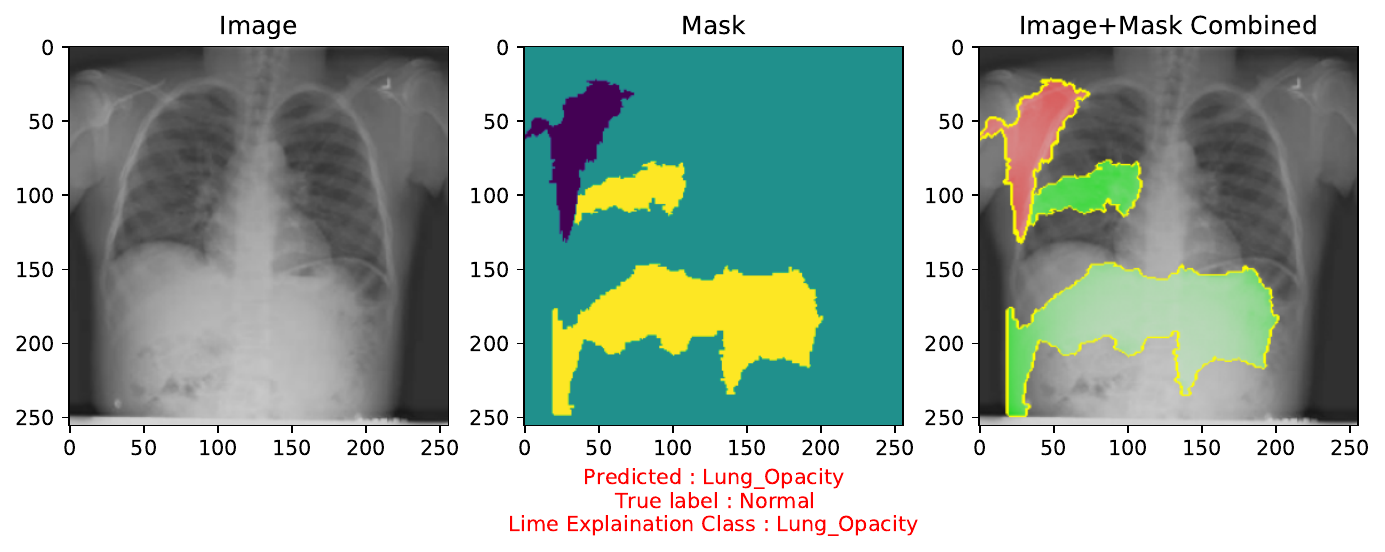}
        \caption{Misclassification Image of Dataset 3}
        \label{Figure 7(c)}
    \end{subfigure}
    \hfill
    \begin{subfigure}[b]{0.48\textwidth}
        \includegraphics[width=\linewidth]{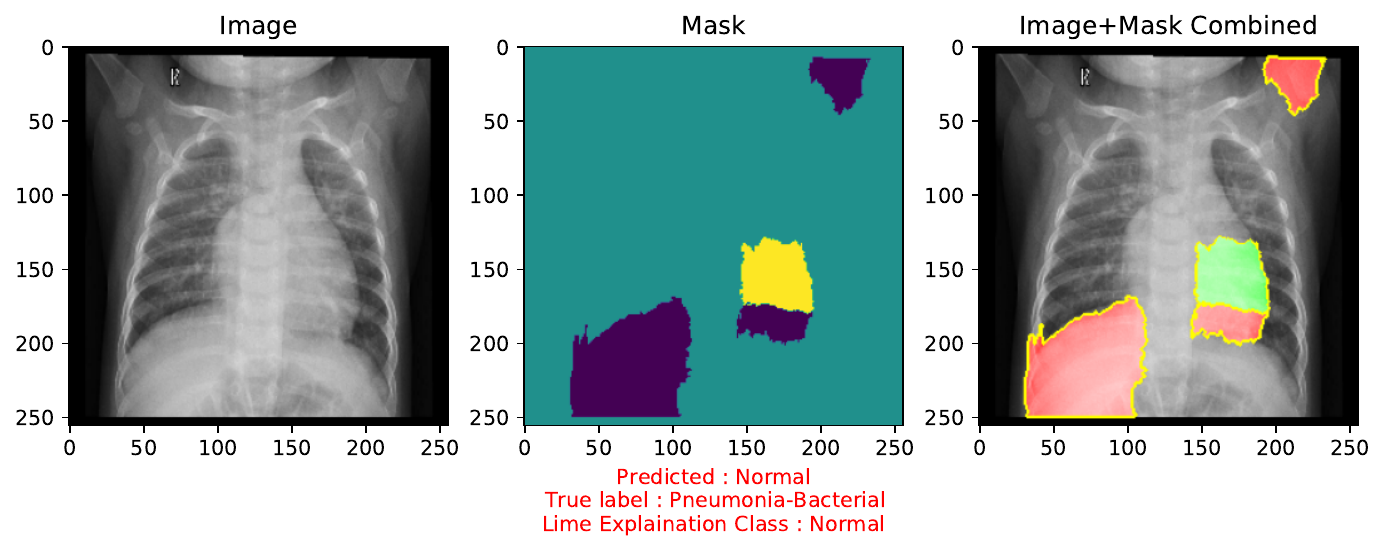}
        \caption{Misclassification Image of Dataset 4}
        \label{Figure 7(d)}
    \end{subfigure}
    
    \vspace{0.5em}
    
    \begin{subfigure}[b]{0.48\textwidth}
        \includegraphics[width=\linewidth]{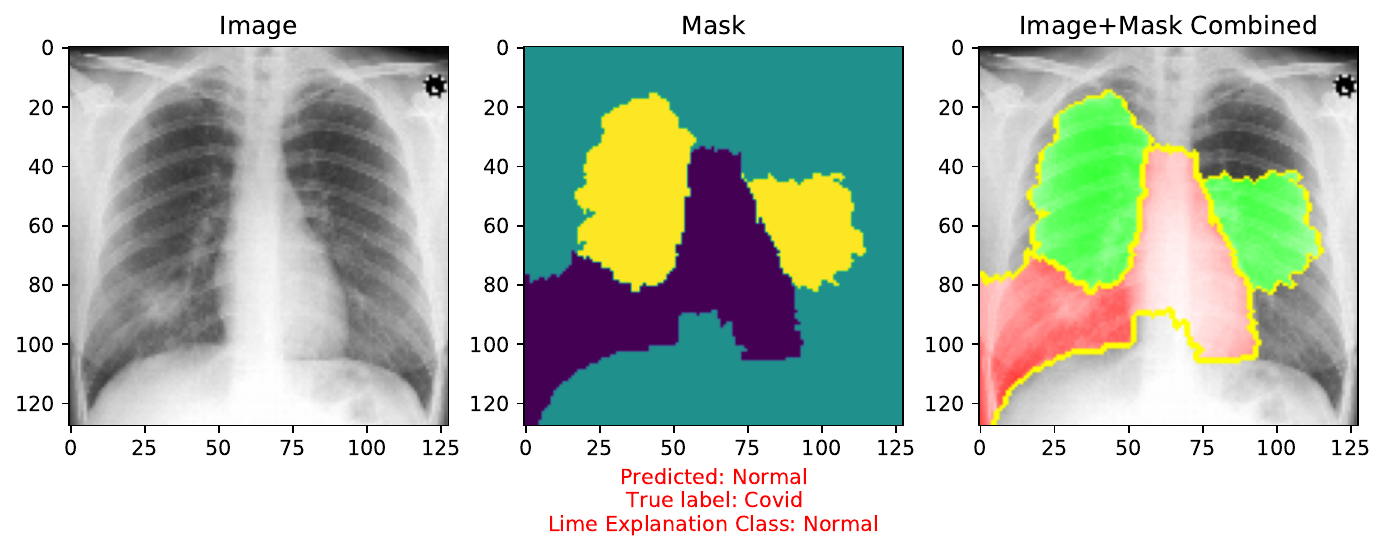}
        \caption{Misclassification Image of Hybrid Dataset}
        \label{Figure 7(e)}
    \end{subfigure}

    \caption{Some Sample Images of Misclassification by the Model}
    \label{Figure 7}
\end{figure}

In \Cref{Figure 7} shows cases where the model misclassified chest X-rays, with LIME explanations providing insights into the model's reasoning. In \Cref{Figure 7(a)} from Dataset 1 \cite{MrinalTyagi}, the model predicts \textit{Normal} when the true label is \textit{Lung Opacity}, with green regions indicating areas the model interprets as \textit{Normal}, leading to the misclassification. The red regions correspond to other classes, but the \textit{Normal} class has a higher probability. In \Cref{Figure 7(b)} from Dataset 2 \cite{https://doi.org/10.17632/m4s2jn3csb.1}, the model incorrectly predicts \textit{COVID} for a case of \textit{Lung Opacity}, with LIME highlighting green regions indicative of \textit{COVID}, which misleads the model. In \Cref{Figure 7(c)} from Dataset 3 \cite{https://doi.org/10.17632/9d55cttn5h.1}, the model predicts \textit{Lung Opacity} instead of the correct \textit{Normal} label, as the LIME explanation shows green areas the model associated with \textit{Lung Opacity}, resulting in the error. For \Cref{Figure 7(d)} from Dataset 4 \cite{https://doi.org/10.17632/9xkhgts2s6.4}, the model predicts \textit{Normal} when the true label is \textit{Pneumonia}, again with green regions supporting the \textit{Normal} prediction, causing the misclassification. Finally, \Cref{Figure 7(e)}  from the hybrid dataset, the model misclassifies a \textit{Covid} case as \textit{Normal}, with LIME explanations showing green areas that the model interprets as \textit{Normal}, leading to the incorrect prediction. From these LIME explanations, we gain valuable insights into the model's black-box decision-making process, enhancing our understanding of its interpretability and identifying misclassification regions.

\section{Limitations \& Future Works}
\label{Limitations and Future Work}
The limitations of this study include regional constraints that may have impacted the generalizability of the findings, as the data primarily reflected specific populations. Additionally, the study employed only one explainable AI (XAI) technique, potentially restricting the depth of interpretability and analysis. Moreover, limitations in the use of transformers and ensemble techniques may have affected the overall performance and robustness of the models. Further research will involve examining larger datasets to enhance the model's robustness and reliability, employing various XAI techniques to explain its decision-making processes, and incorporating data from multiple geographies to mitigate restrictions. Enhancements will concentrate on employing picture segmentation for increased precision, optimizing model architectures, investigating novel techniques like as transformers and ensemble learning, and evaluating the system's applicability in practical medical environments.

\section{Conclusion}
\label{Conclusion}\
\balance
In conclusion, this study demonstrated that integrating regional datasets into a hybrid model significantly enhances the accuracy and generalizability of deep learning models in detecting lung diseases from chest X-rays. The hybrid approach proved effective across diverse datasets, and the use of LIME provided valuable insights into the models' decision-making processes, contributing to a better understanding of how these models classify different lung conditions. The successful application of these techniques underscores the potential for more accurate and interpretable AI-driven medical imaging solutions.

\bibliographystyle{unsrt}  


\bibliography{references.bib}
\end{document}